\documentclass{article}
\usepackage{spconf,amsmath,graphicx}
\usepackage{float}
\usepackage{caption}
\usepackage{subcaption}
\usepackage{nicefrac}
\usepackage{amsmath}
\usepackage{graphicx}
\usepackage{multicol,multirow}
\usepackage{colortbl}
\usepackage{tabularx}
\usepackage{ctable}
\usepackage{cite}
\usepackage{amssymb}
\usepackage{hyperref}
\usepackage{pifont}

\def\pgftextcircled#1{%
        \raisebox{.9pt}{\textcircled{\raisebox{-.9pt}{#1}}}%
}

\title{Can spoofing countermeasure and speaker verification systems \\ be jointly optimised?}
\name{Wanying Ge, Hemlata Tak, Massimiliano Todisco and Nicholas Evans 
\thanks{The first author is supported by the TReSPAsS-ETN project funded by the European Union’s Horizon 2020 research and innovation programme under the Marie Skłodowska-Curie grant agreement No.\ 860813. The second author is supported by the VoicePersonae project funded by the French Agence Nationale de la Recherche (ANR) and the Japan Science and Technology Agency (JST).}}
\address{EURECOM, Sophia Antipolis, France}

\usepackage[pscoord]{eso-pic}
\newcommand{\placetextbox}[3]{
\setbox0=\hbox{#3}
\AddToShipoutPictureFG*{
\put(\LenToUnit{#1\paperwidth},\LenToUnit{#2\paperheight}){\vtop{{\null}\makebox[0pt][c]{#3}}}%
}%
}%

\begin{document}
\maketitle

\begin{abstract}

    Spoofing countermeasure (CM) and automatic speaker verification (ASV) sub-systems can be used in tandem with a backend classifier as a solution to the spoofing aware speaker verification (SASV) task. The two sub-systems are typically trained independently to solve different tasks. While our previous work demonstrated the potential of joint optimisation, it also showed a tendency to over-fit to speakers and a lack of sub-system complementarity. Using only a modest quantity of auxiliary data collected from new speakers, we show that joint optimisation degrades the performance of separate CM and ASV sub-systems, but that it nonetheless improves complementarity, thereby delivering superior SASV performance. Using standard SASV evaluation data and protocols, joint optimisation reduces the equal error rate by 27\% relative to performance obtained using fixed, independently-optimised sub-systems under like-for-like training conditions.
\end{abstract}

\placetextbox{0.5}{0.08}{\fbox{\parbox{\dimexpr\textwidth-2\fboxsep-2\fboxrule\relax}{\footnotesize \centering \copyright  2023 IEEE. Personal use of this material is permitted. Permission from IEEE must be obtained for all other uses, in any current or future media, including reprinting/republishing this material for advertising or promotional purposes, creating new collective works, for resale or redistribution to servers or lists, or reuse of any copyrighted component of this work in other works.}}}

\begin{keywords}
spoofing detection, spoofing countermeasures, speaker verification, joint optimisation
\end{keywords}

\section{Introduction}
\label{sec:intro}

    Solutions to spoofing aware speaker verification (SASV)~\cite{jungsasv2022} typically comprise a pair of essentially-independent sub-systems operating in tandem: a spoofing countermeasure (CM) to verify that the test utterance is bona fide rather than spoofed; an automatic speaker verification (ASV) system to verify whether enrolment and test utterances correspond to the same speaker. Even so, ASV also has potential to detect spoofing attacks. In support of this claim, one can imagine the presentation of poor-quality spoofing attacks, namely artificially generated or manipulated utterances which do not represent well the characteristics of the target speaker. Such poor-quality spoofs may not fool the ASV sub-system and hence be rejected even without an auxiliary CM.

    ASV and CM sub-systems are therefore not independent and there is hence potential for joint optimisation to exploit their complementarity. The CM can be optimised to detect spoofing attacks which {\it will} fool the ASV system, thereby avoiding the unnecessary waste of CM classifier capacity. Despite the appeal of joint optimisation, to the best of our knowledge, no successful solutions have been reported thus far, with all leading solutions combining independently-optimised sub-systems with a backend classifier, e.g.\ using score~\cite{Alenin2022A,dku2022sasv} or embedding~\cite{heo2022IRLab, zhang22b_odyssey} fusion techniques.
    
    Our previous work~\cite{ge2022joint} showed that jointly-optimised SASV solutions can succeed in improving robustness to spoofing but that they have also a tendency to over-fit to the speakers of data used for training. This is hardly surprising; state-of-the-art ASV systems are nowadays trained using data collected from many hundreds of speakers~\cite{voxceleb2} whereas the training partition of the SASV database~\cite{wang2020asvspoof} contains data collected from only 20 speakers. We showed the number of speakers whose data is used for training is a bottleneck and that progress is unlikely to be made without using additional, auxiliary data collected from a larger pool of speakers.

    The work reported in this paper aims to establish the potential and determine whether spoofing countermeasures and speaker verification systems can be jointly optimised. Our hypothesis is that, with data collected from a sufficient speaker population, joint optimisation can deliver superior SASV performance to that obtained using fixed, independently-optimised systems trained under like-for-like training conditions. While joint optimisation might even degrade CM and/or ASV performance, our expectation is that improvements will emerge from more {\it complementary} CM and ASV sub-systems that work in synergy to provide more reliable SASV solutions.

\section{SASV framework}
    
    All work reported in this paper was performed using the same SASV framework used in our previous work~\cite{ge2022joint}. While the framework is the same, there are differences in the training policy in terms of data and trial types. A summary of the framework is presented in this section. The new training policy is described in Section~\ref{sec:database}.
    
\begin{figure}[!t]
         \centering
         \includegraphics[trim=6cm 0 3cm 0,clip,scale=0.5]{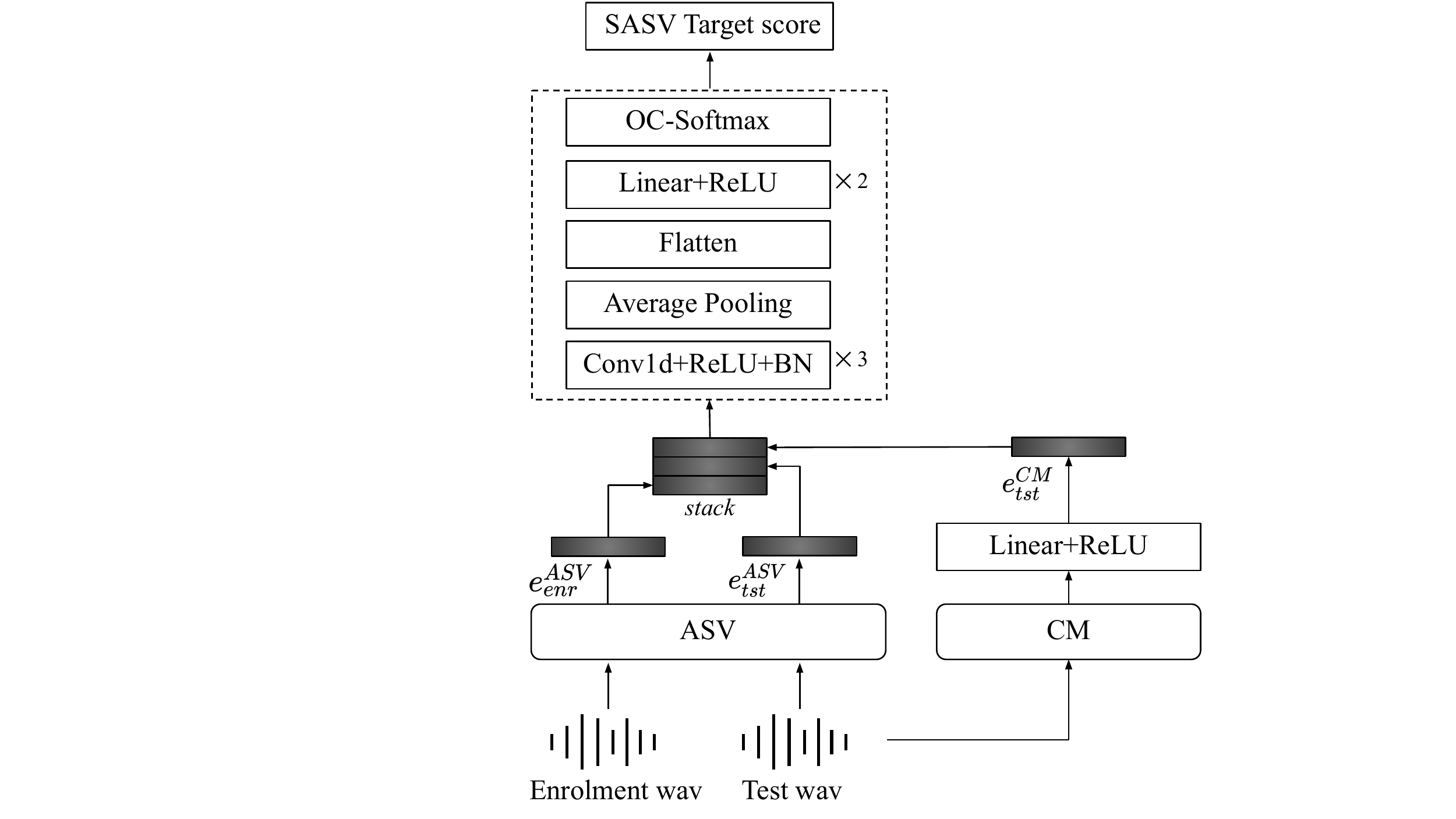}
        \caption{Framework for fixed, independently-optimised and jointly-optimised SASV solutions comprising ASV and CM sub-systems and a backend classifier.}
        \label{fig:sasv}
        \end{figure}

    The architecture is illustrated in Fig.~\ref{fig:sasv} and consists of an ASV sub-system, a CM sub-system and a backend classifier. The ASV sub-system is a ResNet34 model with squeeze-and-excitation (SE) blocks~\cite{hu2018squeeze}, namely the ResNetSE34 model reported in~\cite{heo2020clova}. The input waveform is first decomposed into log Mel-filterbank features. Four convolutional layers with SE blocks and attentive statistics pooling~\cite{okabe2018asp} are used for deep feature extraction and to project the variable length input to a fixed-length embedding vector. A combination of softmax and angular prototypical loss~\cite{wang2017angularloss} is used for optimisation. The ASV sub-system extracts both an enrolment embedding $e^{ASV}_{enr}$ and test embedding $e^{ASV}_{tst}$ from corresponding utterances.
        
    The CM sub-system is the SASV 2022 AASIST~\cite{jung2022aasist} baseline. Raw waveforms are decomposed using a RawNet2-based encoder~\cite{jung20rawnet2,tak2021rawnet2}. Extracted features are integrated using a graph attention network~\cite{tak2021end} before a readout operation and a hidden linear layer are used to generate output scores. Optimisation is performed using a weighted cross-entropy loss function. A linear layer is used to transform representations extracted from the penultimate layer to CM embeddings $e^{CM}_{tst}$ of the same dimension as ASV embeddings.
            
    The backend classifier is a convolutional neural network with an adaptive average pooling layer~\cite{zhang2022flyspeech}. It operates upon the pair of ASV embeddings and the single CM embedding. These are first stacked along a new dimension so that three 1D convolutional layers can be used to capture the variance between speaker representations of both enrolment and test utterances, and also the variance between the ASV and CM embeddings. Deep representations are aggregated using a 1D adaptive average pooling layer and are then mapped to a bona fide, target class score using a pair of linear layers and a one class (OC) softmax layer~\cite{zhang2021oneclass}. A OC-softmax loss function is used to learn higher output scores for the target class.
    
\section{Experimental Setup}
\label{sec:database}

    We describe the set of databases, protocols, assessment and metrics used in this work, together with specific implementation details.

\subsection{Databases}

    Experimental work was performed using three databases. The VoxCeleb2~\cite{voxceleb2} database is used for ASV pre-training. The CM is pre-trained using the training partition of the ASVspoof 2019 Logical Access (LA) database~\cite{wang2020asvspoof}, referred to in the remainder of this paper simply as ASVspoof. The key characteristics of the ASVspoof training partition are detailed in the top half of Table~\ref{tab:database}.  It contains data collected from 20 speakers, a number far less than databases typically used for ASV research~\cite{voxceleb2} and, as we found in previous work~\cite{ge2022joint}, a number that is insufficient to support joint optimisation. All development and evaluation experiments are performed using the corresponding partitions of the ASVspoof database.
    
    The Fake Audio Detection (FAD) database~\cite{ma2022FAD} contains bona fide and spoofed Mandarin-language utterances generated using TTS algorithms. It is used optionally for training the backend classifier or, in the case of joint optimisation, the backend classifier and both CM and ASV sub-systems. The FAD database is sourced from 6 different public databases~\cite{bu_aishell1, shi21c_aishell3, thchs,yang22h_magicdata}. Its non-homogeneous nature makes it ideal as a source of speaker augmentation. To be consistent with the ASVspoof database, we used the data collected from speakers for whom there is both bona fide and spoofed data. The key characteristics of the resulting training partition used in this work are detailed in the lower half of Table~\ref{tab:database}. It contains data collected from 40 speakers, a number twice that of the ASVspoof training partition.

  \begin{table}[!t]
        \caption{Details of the ASVspoof 2019 LA and FAD training partitions used for all experiments reported in this paper.}
        \centerline{
        \resizebox{\columnwidth}{!}{%
        \renewcommand{\arraystretch}{1.15}
        \begin{tabular}{lcccc}
        \hline
        Database & \# of speakers & \# of bona fide & \# of spoofed& \# of attacks\\
        \hline
        ASVspoof train & 20 & 2580 & 22800 & 4 TTS, 2 VC\\
        \hline
        FAD train& 40 & 3200 & 25600 & 8 TTS\\
        \hline
        \end{tabular}}
        }
        \label{tab:database}
        \end{table}    
      
\subsection{Protocols}

    We used SASV~\cite{jungsasv2022} protocols for the ASVspoof database for all work reported in this paper, modifying only the training protocol to incorporate optional FAD data. For fixed optimisation, pre-trained sub-systems are used without additional optimisation. The backend is optimised according to one of three different training conditions using either: ASVspoof data alone; ASVspoof + FAD data; ASVspoof + FAD bona fide data only. The last training condition is included since we aim to reduce the over-fitting of the ASV sub-system which is usually trained with bona fide data only. In doing so we avoid, to the extent possible, domain mismatch in terms of spoofing attacks (which is not the goal of this work); domain mismatch in the bona fide data nonetheless remains. Only for the jointly-optimised system are all three components trained simultaneously through back propagation.

    Training is performed using different combinations of trial pairs including a bona fide target enrolment utterance in addition to a test utterance. The latter can be: \pgftextcircled{1} a bona fide target speaker utterance; \pgftextcircled{2} a bona fide non-target speaker utterance; \pgftextcircled{3} a spoofed target speaker utterance. The proportion of each trial type for fixed and jointly-optimised systems is illustrated in Table~\ref{tab:classes}. For joint optimisation we found the use of a fourth trial type involving \pgftextcircled{4} a spoofed non-target speaker utterance to be beneficial, but found no improvement for our fixed system. The full SASV system should accept only type \pgftextcircled{1} trials. All others should be rejected, even though type \pgftextcircled{2} trials are within the positive class for the CM.\footnote{By convention, the CM produces higher scores for bona fide inputs and lower scores for spoofed inputs.} The conflict between SASV and CM classes then results in degraded CM performance through back propagation. The introduction of type \pgftextcircled{4} trials acts to compensate for this behaviour since they are among the negative class for SASV and both sub-systems. Finally, the development and evaluation protocols are the standard SASV protocols reported in~\cite{jungsasv2022,shim2022baseline}.
        
\subsection{Assessment and metrics}

    We use the output scores from the backend OC-softmax layer to assess the performance of both fixed and jointly-optimised systems. In order to analyse impacts of joint optimisation and complementarity, we also assess performance at the sub-system level. ASV sub-system scores are the cosine distance between enrolment and test utterance embeddings. CM sub-system scores are the AASIST outputs for the bona fide class and test utterance only. Three different SASV metrics, all equal error rate (EER) estimates, are used to assess speaker verification performance (SV-EER) involving a set of \pgftextcircled{1} and \pgftextcircled{2} trials, spoofing detection performance (SPF-EER) involving a set of \pgftextcircled{1} and \pgftextcircled{3} trials and spoofing-aware speaker verification (SASV-EER) performance involving the full set of \pgftextcircled{1}, \pgftextcircled{2} and \pgftextcircled{3} trials. The remaining trial type \pgftextcircled{4} is not used for assessment, neither for development, nor for evaluation.
        
        \begin{table}[!t]
        \caption{SASV training pairs and the corresponding proportions for fixed and joint optimisation.}
        \centerline{
        \resizebox{\linewidth}{!}{
        \renewcommand{\arraystretch}{1.1}
        \begin{tabular}{cccc}
        \hline
        Pair & Test utterance  & Prop. in Fixed & Prop. in Joint\\
        \hline
        \multirow{1}{*}{\pgftextcircled{1}} & Bona fide, Target spk & \multirow{1}{*}{50\%} & \multirow{1}{*}{25\%}  \\
        \hline
        \multirow{1}{*}{\pgftextcircled{2}} & Bona fide, Non-target spk & \multirow{1}{*}{25\%} & \multirow{1}{*}{25\%}  \\
        \hline
        \multirow{1}{*}{\pgftextcircled{3}} & Spoofed, Target spk & \multirow{1}{*}{25\%} & \multirow{1}{*}{25\%}  \\
        \hline
        \multirow{1}{*}{\pgftextcircled{4}} & Spoofed, Non-target spk & \multirow{1}{*}{None} & \multirow{1}{*}{25\%}  \\
        \hline
        \end{tabular}}
        }
        \label{tab:classes}
        \end{table}
        
\subsection{Implementation}
        
    All trainable network parameters are updated for 20 epochs with an initial learning rate of 5e-5. The batch size is set to 20 for both systems. Model selection is made according to the lowest SASV-EER estimate for the development partition. Systems performance is estimated from the average of 5 independent runs each with a different random seed. All experiments were performed on a single NVIDIA GeForce RTX 3090 GPU. Results are reproducible with the same set of random seeds and GPU environment using the implementation available online.\footnote{\url{https://github.com/eurecom-asp/sasv-joint-optimisation}}
        
\section{Results}

    \begin{table*}[!t]
    \caption{Averaged results for pre-trained, jointly-optimised systems for SASV 2022 evaluation partitions. Results for ASV and CM sub-systems under fixed configuration are same since the networks are identical. The corresponding results for full system are different because their backend classifiers are trained using different databases. Results in \textbf{boldface} indicate better performance for jointly-optimised systems than for corresponding fixed, independently-optimised systems.}
    \vspace{-6mm}
    \begin{center}
    \resizebox{\textwidth}{!}{
    \renewcommand{\arraystretch}{1.2}
    \begin{tabular}{lcccccccccc}
    \toprule
        \multirow{2}{*}{\textbf{Training data}} & \multirow{2}{*}{\textbf{Configuration}} & \multicolumn{3}{c}{\textbf{Full system}} & \multicolumn{3}{c}{\textbf{ASV sub-system}}& \multicolumn{3}{c}{\textbf{CM sub-system}}  \\
        \cmidrule(lr){3-5} \cmidrule(lr){6-8} \cmidrule(lr){9-11} 
        \multicolumn{1}{c}{}  & \multicolumn{1}{c}{} & SASV-EER & SV-EER & SPF-EER & SASV-EER & SV-EER & SPF-EER & SASV-EER & SV-EER & SPF-EER  \\ 
        \hline
        \multirow{2}{*}{ASVspoof} & Fixed & 1.15 &  1.27 & 1.08 & 19.70 & 1.27 & 25.75 & 24.50 & 49.01 & 0.65\\
         & Joint &1.49 & 2.34 & \textbf{0.80} & \textbf{13.47} & {1.84} & \textbf{17.43} & \textbf{23.65} & \textbf{46.80} & {1.02} \\
         \hline
         \multicolumn{1}{l}{ASVspoof +} & \multicolumn{1}{c}{Fixed} & 1.52 & {1.85}  & {1.27}  & 19.70 & {1.27} & {25.75} & {24.50} & {49.01} & {0.65}\\
         \multicolumn{1}{l}{FAD} & \multicolumn{1}{c}{Joint} &  1.74 & {2.66} & \textbf{1.09} & \textbf{8.57} & {1.73} & \textbf{10.82} & \textbf{22.83} & \textbf{46.37} & {1.81} \\
         \hline
         \multicolumn{1}{l}{ASVspoof +} & \multicolumn{1}{c}{Fixed} &1.72 & {1.47} & {1.84} & 19.70 &  {1.27} & {25.75} & {24.50} & {49.01} &  {0.65}\\
         \multicolumn{1}{l}{FAD bona fide only} & \multicolumn{1}{c}{Joint} & \textbf{1.26} & {1.77} & \textbf{0.83} & \textbf{8.58} & {1.49} & \textbf{10.60} & {24.66} & \textbf{47.99} & {1.35} \\
    \bottomrule
    \end{tabular}
    }
    \end{center}
    \vspace{-6mm}
    \label{tab:sasv_large}
    \end{table*}

    Results are shown in Table~\ref{tab:sasv_large}. It shows SASV-EER, SV-EER, and SPF-EER estimates for the full system (left), the ASV sub-system (middle) and CM sub-system (right) when training is performed using ASVspoof data alone (top), ASVspoof and FAD data (middle) and then ASVspoof and only FAD bona fide data (bottom). In the following we outline a number of principle observations and supporting results.
    
    {\bfseries CM performance} -- SPF-EER estimates for the CM sub-system in the last column of Table~\ref{tab:sasv_large} show that the lowest SPF-EER (0.65\%) is obtained using a fixed system trained using only ASVspoof data. The degradations in CM performance (1.02\%, 1.81\% and 1.35\%) stem from the CM being optimised to detect only attacks that are successful in fooling the ASV system. While there is no need for the CM to detect less potent attacks, but since performance estimates are still made using the full set, the SPF-EER increases. The worst result is for the ASVspoof + FAD training condition, and is the result of differences between spoofing attacks in the ASVspoof and FAD databases. SPF-EER estimates for the ASV sub-system (25.75\%) show that even the fixed system can detect spoofing attacks, but that joint optimisation acts to improve performance (17.43\%, 10.82\% and 10.60\%). SPF-EER results for the full system show the complementarity of the ASV and CM sub-systems in improving overall robustness to spoofing for all three training conditions.
    
    {\bfseries ASV performance} -- SV-EER estimates for the CM sub-system shown in the penultimate column of Table~\ref{tab:sasv_large} show evidence of {\it speaker-awareness}. Without over-fitting to speakers, the CM sub-system should achieve an SV-EER of 50\%. Results for jointly-optimised systems (46.8\%, 46.37\% and 47.99\%) deviate further than the result for the fixed system (49.01\%). Corresponding SV-EER results for the ASV sub-system show that joint optimisation acts to degrade performance, with the best result being achieved using the fixed system (1.27\%). This translates to worse ASV performance for the full system; SV-EER results for all jointly-optimised systems (2.34\%, 2.66\% and 1.77\%) are all worse than those for fixed systems (1.27\%, 1.85\% and 1.47\%). Again, the worst result is for the ASVspoof + FAD training condition.
    
    {\bfseries SASV performance} -- Joint optimisation makes little difference to the SASV-EER for the CM sub-system (24.5\% vs.\ 23.65\%, 22.83\% and 24.66\%). This is not surprising, since the CM sub-system operates only upon the test utterance. Nonetheless, joint optimisation improves ASV sub-system performance, with results for all jointly-optimised systems (13.47\%, 8.57\% and 8.58\%) being better than the fixed system (19.7\%). Even if there is little difference in CM results, but substantial improvements to ASV results, this does not translate to better performance for the full system in the case of the ASVspoof (1.15\% vs.\ 1.49\%) and ASVspoof + FAD (1.52\% vs.\ 1.74\%) training conditions. Only for the ASVspoof + FAD bona fide condition is joint optimisation beneficial (1.72\% vs. 1.26\%).  
    
\section{Discussion}
    
    The lowest SASV-EER comes from the fixed system trained using ASVspoof data alone. This result may cast doubt upon the claimed merit since, even with more training data, the jointly-optimised system does not outperform the fixed counterpart. However, while the jointly-optimised system trained using ASVspoof + FAD bona fide only data does give performance that is behind that of the fixed system trained with ASVspoof data only, the use of different training data makes for an unfair comparison. The FAD training data is out-of-domain in the form of speech data in a different language and recording conditions, among other differences. In this sense, comparisons should be made only between results for the {\it same} training conditions, i.e.\ results for ASVspoof {\it or} ASVspoof + FAD bona fide only training conditions. In this sense, lower EERs observed across different training conditions are not an indication of joint optimisation not working, but are the result of domain mismatch which is not tackled in this work. Domain mismatch is a penalty which likely degrades results for training conditions that use FAD data. The use of domain-matched training data, or domain adaptation, may than show the full potential and even better results.
    
    Results for the ASVspoof and FAD bona fide only training condition show a 27\% relative reduction in the SASV-EER under like-for-like training conditions. It comes as a result of using data collected from a greater number of speakers and jointly-optimised CM and ASV sub-systems with {\it worse} performance than their independently-optimised counterparts. Even so, the resulting sub-systems are more {\it complementary} leading to better overall reliability (SASV-EER).
    
\section{Conclusions and further work}

    Results presented in this paper add to the evidence that joint optimisation has potential to better exploit the synergy between spoofing countermeasures and speaker verification sub-systems so that they function cooperatively as a more reliable solution to spoofing aware speaker verification (SASV). Interestingly, results show that joint optimisation degrades the performance of each sub-system but that it improves their complementarity. Joint optimisation improves SASV performance by 27\% relative to its fixed, independently-optimised counterpart under like-for-like training conditions. This result is obtained despite using a training database of the same size to jointly-optimise a considerably more complex system.

    While other approaches to joint optimisation might make better use of data collected from fewer speakers, our solution is only successful when using a modest quantity of auxiliary data collected from new speakers. Given that state-of-the-art ASV solutions are trained with data collected from many hundreds of speakers, it seems wise for future ASVspoof and SASV challenges to collect and make available training data collected from far more speakers than in past editions. The domain robustness problem persists; even if we can learn speaker variation in a joint optimisation framework using bona fide data sourced from a different database, it does not translate to reliable detection for spoofing attacks in the same database. Research in domain robustness is hence a priority. Other directions include the investigation of joint optimisation strategies to reduce CM over-fitting; the CM shows signs of speaker awareness. Speaker-dependent spoofing detection might also be an interesting avenue for future research.
    
\vfill\pagebreak
\bibliographystyle{IEEEbib}
\bibliography{refs}

\end{document}